\documentclass[twocolumn]{aastex63}

\usepackage{amsmath}

\hypersetup{linkcolor=red,citecolor=blue,filecolor=green,urlcolor=magenta}

\begin{document}

\shorttitle{Yellow Post-AGB Stars}
\shortauthors{Ngeow et al.}

\title{Zwicky Transient Facility and Globular Clusters: Calibration of the $gr$-Band Absolute Magnitudes for the Yellow Post-Asymptotic-Giant-Branch Stars}

\correspondingauthor{C.-C. Ngeow}
\email{cngeow@astro.ncu.edu.tw}

\author[0000-0001-8771-7554]{Chow-Choong Ngeow}
\affil{Graduate Institute of Astronomy, National Central University, 300 Jhongda Road, 32001 Jhongli, Taiwan}

\author[0000-0001-6147-3360]{Anupam Bhardwaj}
\affil{INAF-Osservatorio astronomico di Capodimonte, Via Moiariello 16, 80131 Napoli, Italy}

\author{Daniel Reiley}
\affil{Caltech Optical Observatories, California Institute of Technology, Pasadena, CA  91125, USA}

\author[0000-0003-2451-5482]{Russ R. Laher}
\affiliation{IPAC, California Institute of Technology, 1200 E. California Blvd, Pasadena, CA 91125, USA}

\author[0000-0003-1227-3738]{Josiah Purdum}
\affiliation{Caltech Optical Observatories, California Institute of Technology, Pasadena, CA 91125, USA} 

\author[0000-0001-7648-4142]{Ben Rusholme}
\affiliation{IPAC, California Institute of Technology, 1200 E. California Blvd, Pasadena, CA 91125, USA}

\begin{abstract}

  We present the first absolute calibration for the yellow post-asymptotic-giant-branch (PAGB) stars in the $g$- and $r$-band based on time-series observations from the Zwicky Transient Facility. These absolute magnitudes were calibrated using four yellow PAGB stars (one non-varying star and three Type II Cepheids) located in the globular clusters. We provide two calibrations of the $gr$-band absolute magnitudes for the yellow PAGB stars, by using an arithmetic mean and a linear regression. We demonstrate that the linear regression provides a better fit to the $g$-band absolute magnitudes for the yellow PAGB stars. These calibrated $gr$-band absolute magnitudes have a potential to be used as population II distance indicators in the era of time-domain synoptic sky surveys.

\end{abstract}


\section{Introduction}\label{sec1}

Post-asymptotic-giant-branch (hereafter PAGB) stars are the low- to intermediate-mass stars at their final and short-lived stage of stellar evolution before entering the planetary nebulae phase. PAGB stars evolve with nearly constant luminosity but with increasing effective temperature on the Hertzsprung-Russell (H-R) diagram. During such evolution, these PAGB stars become Type II Cepheids, or more specifically the RV Tauri (RV Tau) stars \citep[for example, see][]{jura1986,alcock1998}\footnote{RV Tau stars, however, could be a mixed of both PAGB stars and post red-giant-branch stars \citep[for example, see][and reference therein]{manick2018,giridhar2020}.}, when they cross the instability strip on the H-R diagram.  

The idea of using PAGB stars as population II distance indicators has been proposed in the past \citep{bond1997}. This is because PAGB stars have the largest luminosity for the low- to intermediate-mass stars throughout their lifetime, with $M_\mathrm{bol}\sim -3.38$~mag based on ten yellow, or intermediate-temperature, PAGB stars located in seven globular clusters \citep{ciardullo2022}. Since half of them are Type II Cepheids, \citet{ciardullo2022} calibrated the Johnson $V$-band absolute magnitude of yellow PAGB stars using the nonvariable stars with colors in the range of $0.0\lesssim (B-V) \lesssim 0.5$~mag, i.e. blueward of the instability strip, and obtained $M_V=-3.37\pm0.05$~mag. The reasons, or advantages, for choosing the blue colors have been extensively discussed in \citet{ciardullo2022}. Obviously, selecting the nonvarying yellow PAGB stars (blueward of instability strip) has an advantage on reducing the observing time, because in general RV Tau stars have pulsation periods longer than $\sim 20$~days. As pulsating stars, RV Tau stars also obey a period-luminosity-color relation, implying a color-magnitude relation (at fixed period) on the color-magnitude diagram (CMD).

Given that the $ugrizy$ filters, or a subset of them, are increasingly popular among various synoptic sky surveys, with a representative example of the Vera C. Rubin Observatory's Legacy Survey of Space and Time \citep[LSST,][]{lsst2019}, it is desirable to calibrate the absolute magnitudes of the yellow PAGB stars in these filters in addition to the $V$-band. Similar to the work of \citet{ciardullo2022} in $V$-band, our goal is to calibrate the $gr$-band absolute magnitudes for the yellow PAGB stars located in the globular clusters by using the photometric data obtained from the Zwicky Transient Facility \citep[ZTF,][]{bellm2019,gra19,mas19,dec20}. Out of the ten yellow PAGB stars listed in \citet{ciardullo2022}, four of them are located in the southern globular clusters ($\omega$~Centauri and NGC~5986), which are outside the footprint of ZTF. Hence, our calibration was done on the remaining non-variable yellow PAGB stars and the Type II Cepheids.

\section{ZTF Mean Magnitudes} \label{sec2}

\begin{figure}
  \epsscale{1.1}
  \plotone{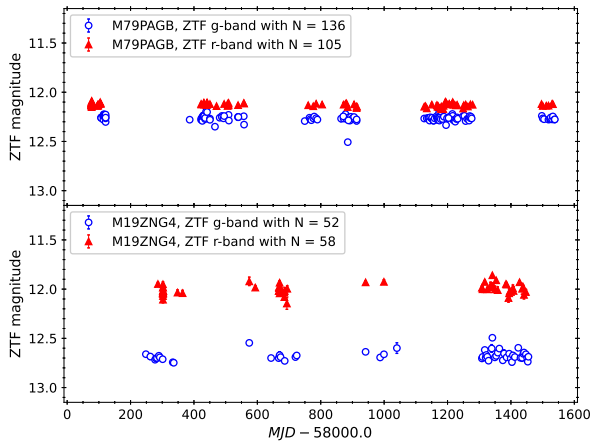}
  \caption{ZTF $gr$-band light curves for M79 PAGB (top panel) and M19 ZNG4 (bottom panel), after excluding the data points with non-zero {\tt flags} and {\tt infobits}. The remaining numbers of data points are listed as $N$.}
  \label{fig_lc}
\end{figure}

ZTF is a 47-squared-degree wide-field synoptic northern sky survey utilizing the Palomar 48-inch Samuel Oschin Telescope located in southern California, USA. The majority of the ZTF observations were conducted with the customized $g$ and $r$ filters, with additional $i$-band observations from the partner surveys.\footnote{ZTF observations were divided into public surveys, partner surveys, and Caltech (California Institute of Technology) surveys. For more details, see \citet{bellm2019}.} All of the ZTF images were processed through a dedicated reduction pipeline \citep{mas19}, and the ZTF photometry was calibrated to the Pan-STARRS1 \citep[Panoramic Survey Telescope and Rapid Response System 1,][]{chambers2016,magnier2020} AB magnitude system.

As in \citet{ngeow2022}, ZTF $gr$-band light curves for the two non-variable yellow PAGB stars, M19 ZNG4 and M79 PAGB, were extracted from the ZTF PSF (point-spread function) catalogs. No ZTF $i$-band data were available for these two targets in the ZTF Public Data Release 10 (DR10) and the partner surveys data until 2022 March 31. After excluding data points with non-zero {\tt flags} and {\tt infobits},\footnote{For the definitions and meanings of {\tt flags} and {\tt infobits}, see ``The ZTF Science Data System (ZSDS) Explanatory Supplement'' document, available at \url{https://irsa.ipac.caltech.edu/data/ZTF/docs/ztf_explanatory_supplement.pdf}.} ZTF light curves for these two targets are presented in Figure \ref{fig_lc}. As in \citet{bond2016}, we do not detect obvious variability for M79 PAGB at $\sim0.03$~mag level (the dispersion of the light curves). In case of M19 ZNG4, \citet{bond2021} suspected this star does not exhibit variability based on two observations separated by one night (plus earlier low-quality data). The ZTF light curves confirmed this target does not exhibit obvious variability at $\sim0.05$~mag level. Hence, we obtained straight weighted mean magnitudes in the $gr$-band for these two yellow PAGB stars, as reported in Table \ref{tab_absmag}.\footnote{The $g$-band mean magnitude for M79 PAGB remained unchanged whether the ``outlier'' point at $g\sim12.5$~mag was excluded or not.}

Among the four Type II Cepheids listed in \citet{ciardullo2022} that are observable with ZTF, three of them (V11 in M2, V42 and V84 in M5) have $gr$-band mean magnitudes derived in \citet{ngeow2022}, and they are given in Table \ref{tab_absmag}. The Type II Cepheid V17 in M28 was excluded in \citet{ngeow2022} due to blending.

\begin{deluxetable}{lcccc}
  \tabletypesize{\scriptsize}
  \tablecaption{Apparent and Absolute Magnitudes of the Targeted Stars\label{tab_absmag}}
  \tablewidth{0pt}
  \tablehead{
    \colhead{Target} &
    \colhead{$\langle g \rangle$} &
    \colhead{$\langle r \rangle$} &
    \colhead{$M_g$} &
    \colhead{$M_r$} 
  }
  \startdata  
  M19 ZNG4 & $12.676$ & $12.014$ & $-3.647\pm0.046$ & $-3.869\pm0.044$ \\
  M79 PAGB & $12.265$ & $12.128$ & $-3.318\pm0.030$ & $-3.455\pm0.030$ \\
  \hline
  M5 V42\tablenotemark{a}   & $11.457$ & $11.123$ & $-3.229\pm0.018$ & $-3.482\pm0.018$ \\
  M5 V84\tablenotemark{a}   & $11.626$ & $11.231$ & $-3.138\pm0.019$ & $-3.432\pm0.019$ \\
  M2 V11\tablenotemark{a}   & $12.300$ & $11.933$ & $-3.039\pm0.021$ & $-3.406\pm0.021$ \\
  \enddata
  \tablenotetext{a}{Values of $\langle g \rangle$ and $\langle r \rangle$ were adopted from \citet{ngeow2022}.}
\end{deluxetable}

\section{Analysis and Results} \label{sec3}

\begin{figure*}
  \epsscale{1.1}
  \plotone{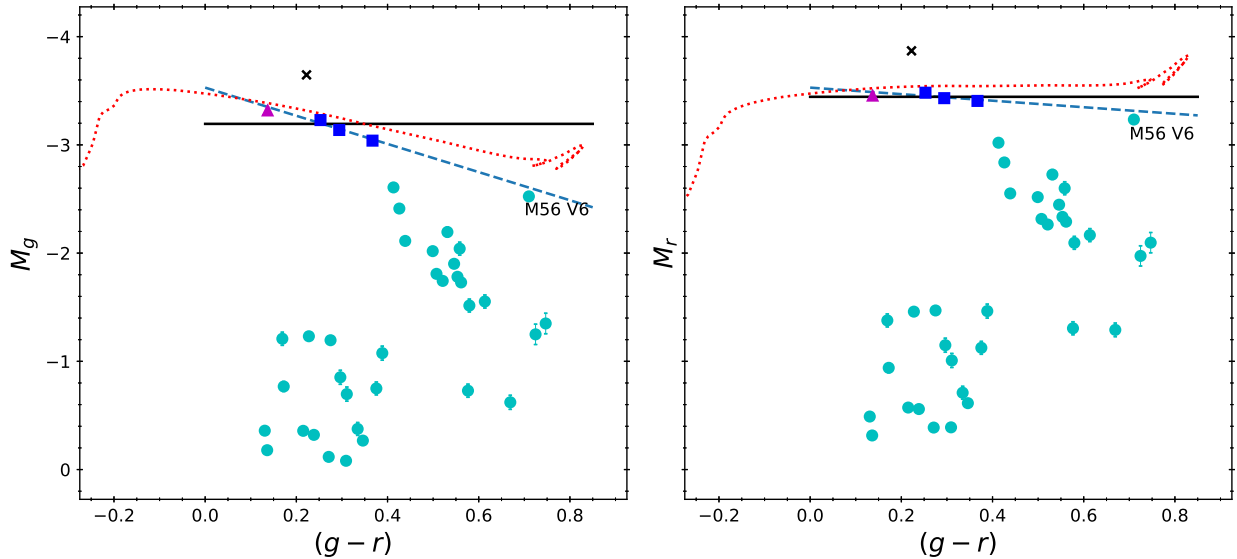}
  \caption{Extinction-corrected CMD for M79 PAGB (magenta triangles), the three Type II Cepheids (blue squares), and other Type II Cepheids compiled in \citet[][cyan circles]{ngeow2022}. The crosses marked the discarded yellow PAGB star, M19 ZNG4. The solid lines represent the mean absolute magnitudes as given in equation (1) and (2), while the dashed lines are the fitted linear regressions given equation (3) and (4), extrapolated to redder colors. The dotted lines are the theoretical PAGB evolution track adopted from \citet{moehler2019}, with a metallicity of $-2.3$~dex and a zero-age horizontal-branch mass of $0.55\ M_\odot$, converted to the observed planes using the {\tt PARSEC Bolometric Correction} online tool. Note that the track is used for illustration purpose and not for fitting the data. Error bars on the majority of the data points are smaller than the size of the symbols.}
  \label{fig_cmd}
\end{figure*}

Apparent mean magnitudes of the five targets listed in Table \ref{tab_absmag} were converted to absolute magnitudes by adopting the distances to their host globular clusters from \citet{baumgardt2021}, with extinction corrections based on the the {\tt Bayerstar2019} 3D reddening map \citep{green2019}. For more details on the extinction corrections, see \citet{ngeow2022}. Errors on the absolute magnitudes were based on the propagated errors on the apparent mean magnitudes (ranging from $\sim 0.002$~mag to $\sim 0.008$~mag), the errors on the distance, and the errors returned from the {\tt Bayerstar2019} 3D reddening map. The final $gr$-band absolute magnitudes are listed in the last two columns of Table \ref{tab_absmag}.

The absolute magnitudes for M19 ZNG4 seem to be brighter than other targets listed in Table \ref{tab_absmag}, which hints at an issue of blending. Indeed, if we transformed the $gr$-band apparent magnitudes to the $V$-band via the transformation provided in \citet{tonry2012}, we obtained $V=12.333\pm0.014$~mag. This $V$-band magnitude is brighter than the value of $12.512\pm0.006$~mag given in \citet{bond2021}, suggesting additional fluxes were included presumably due to blending. In contrast, the transformed $V$-band apparent magnitude for M79 PAGB is $12.197\pm0.012$~mag, in good agreement with the value found in \citet[][$12.203\pm0.008$~mag]{bond2016}. Hence, we discarded M19 ZNG4 for the calibration of the $gr$-band absolute magnitudes for the yellow PAGB stars.

Based on M79 PAGB and the three Type II Cepheids, the weighted means of the $gr$-band absolute magnitudes are

\begin{eqnarray}
  M_g & = & -3.19 \pm 0.14\ \ \mathrm{mag}, \\
  M_r & = & -3.44 \pm 0.04\ \ \mathrm{mag}. 
\end{eqnarray}

\noindent Errors on these absolute magnitudes were calculated based on small number statistics \citep[][p. 202]{dean1951,keeping1962}, as there are only four stars in the sample. These mean absolute magnitudes are marked as solid lines in Figure \ref{fig_cmd}. The $g$-band absolute magnitude exhibits a larger error due to its larger color-dependency as shown in the left panel of Figure \ref{fig_cmd}. We have overlaid a PAGB evolutionary track, selected from models constructed in \citet{mb2016} and \citet{moehler2019}, in Figure \ref{fig_cmd}. The theoretical luminosities and effective temperatures along this track have been converted to the $gr$-band absolute magnitudes using an online tool {\tt PARSEC Bolometric Correction} \citep{chen2019}.\footnote{\url{http://stev.oapd.inaf.it/YBC/}} The PAGB evolution track shows a larger gradient on the $g$-band CMD than the $r$-band CMD, explaining the larger error on equation (1). Therefore, we fit a linear regression to these four stars and yields:\footnote{Due to small number of data points, errors on the regression coefficients given in equation (3) and (4) might be under-estimated. Using alternative pair bootstrap regression method (note that this method might not valid for small number of data points), we found that errors on these regression coefficients could be $\sim2$ to $\sim10$~times larger. Therefore, when applying equation (3) and/or (4) in distance scale applications, extra care has to be taken into account for estimating the errors in the derived distances. For a throughout discussion on errors propagation for distance scale estimation, see the appendix in \citet{feigelson1992}.}

\begin{eqnarray}
  M_g & = & 1.30[\pm0.15](g-r) - 3.53[\pm0.04], \\
  M_r & = & 0.30[\pm0.15](g-r) - 3.53[\pm0.04].
\end{eqnarray}

\noindent Both linear regressions give the same standard deviation of $0.03$~mag. As demonstrated in Figure \ref{fig_cmd}, in the $g$-band these four stars are better described with a linear regression than adopting a mean value. In contrast, the PAGB evolutionary track is almost a constant in the $r$-band, explaining the much smaller error in the $r$-band mean absolute magnitude, i.e. equation (2), than the $g$-band. 

\begin{figure}
  \epsscale{1.1}
  \plotone{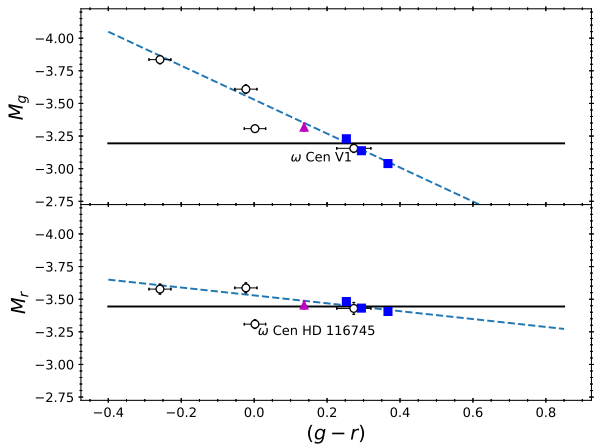}
  \caption{Similar to Figure \ref{fig_cmd}, but with additional four yellow PAGB stars in the southern globular clusters (open circles, see text for details). Note that the dispersions of the photometric transformations as given in \citet{tonry2012} dominate the errors on the $(g-r)$ colors for these four southern yellow PAGB stars. V1 in $\omega$~Centauri ($\omega$~Cen) is a Type II Cepheid.}
  \label{fig_zoom}
\end{figure}

We tested our results using the four yellow PAGB stars located in the southern globular clusters $\omega$~Centauri and NGC~5986. The $BV$-band photometry of these four stars, adopted from \citet[][for V1 in $\omega$~Centauri]{braga2020} and \citet[][for HD 116745 in $\omega$~Centauri and the two PAGB stars in NGC~5986]{davis2022}, were transformed to $gr$-band using the relations given in \citet{tonry2012}. Since  the {\tt Bayerstar2019} 3D reddening map did not cover the sky south of $-30^\circ$ declination, we queried the \citet{sfd1998} extinction map to correct the extinctions for these four southern yellow PAGB stars \citep[see][on the procedure of extinction corrections]{ngeow2022}. Figure \ref{fig_zoom} shows the positions of these four southern yellow PAGB stars on the extinction-corrected CMD as open circles. Except for HD 116745 in $\omega$~Centauri, other three southern yellow PAGB stars are closer to the linear regressions given in equation (3) and (4) extrapolated to bluer colors, supporting the use of linear regressions for calibrating the absolute magnitudes of yellow PAGB stars, especially in the $g$-band.

A Type II Cepheid, V6 in M56, appears to be located near the extrapolated linear regressions given in equation (3) and (4) to the redder colors. This Type II Cepheid is indeed a PAGB star \citep{davis2022} with a redder color of $(g-r)=0.710$~mag. Its $M_V=-2.95$~mag is marginally fainter than the $M_V=-3.0$~mag cut when selecting the PAGB stars as calibrators for the distance indicators \citep{ciardullo2022}. This star has a larger deviation from the mean $gr$-band absolute magnitudes (the solid lines in Figure \ref{fig_cmd}), and would be treated as outlier. Nevertheless, this Type II Cepheid is located redward of the instability strip, hence it is not preferable to be included as a calibrator \citep{ciardullo2022}.

\section{Discussion and Conclusions} \label{sec4}

In this work, we have derived the $gr$-band absolute magnitudes for yellow PAGB stars that can be used as population II distance indicators. Both the mean absolute magnitudes and color-dependent linear regressions were derived based on four yellow PAGB stars located in the globular clusters, by using the homogeneous ZTF data and adopting the same sources for the distance to the globular clusters and reddening corrections. In this way, we minimize the possible systematic errors arising from inhomogeneous data sets.

Since there are only four stars (one non-varying star and three Type II Cepheids) in our sample as calibrators, we were forced to include the variable Type II Cepheids in the calibration processes \citep[in contrast to][]{ciardullo2022}. Nevertheless, Figure \ref{fig_cmd} demonstrates that the Type II Cepheids can be included together with the non-varying yellow PAGB star to increase the sample size. Certainly, searching for the non-varying yellow PAGB stars in distant galaxies only requires single-shot observations, which is advantageous for expensive or competitive observations such as using the {\it Hubble Space Telescope} \citep{ciardullo2022}. On the other hand, long-term synoptic sky surveys such as LSST will naturally provide time-series data to search for the long period Type II Cepheids, as long as these Type II Cepheids or RV Tau stars could be identified as the low-surface-gravity yellow PAGB stars.\footnote{Recall that not all RV Tau stars are PAGB stars. How to identify PAGB stars using conventional broad-band photometry is beyond the scope of this paper.} This would increase the number of suitable yellow PAGB stars to serve as distance indicators, hence reducing the statistical errors when deriving the distances to their host galaxies.

\acknowledgments

We thank M. M. Miller Bertolami for sharing the theoretical evolutionary tracks, and the useful comments from an anonymous referee that improved the manuscript. We are thankful for funding from the Ministry of Science and Technology (Taiwan) under the contracts 107-2119-M-008-014-MY2, 107-2119-M-008-012, 108-2628-M-007-005-RSP and 109-2112-M-008-014-MY3.

Based on observations obtained with the Samuel Oschin Telescope 48-inch Telescope at the Palomar Observatory as part of the Zwicky Transient Facility project. ZTF is supported by the National Science Foundation under Grants No. AST-1440341 and AST-2034437 and a collaboration including current partners Caltech, IPAC, the Weizmann Institute of Science, the Oskar Klein Center at Stockholm University, the University of Maryland, Deutsches Elektronen-Synchrotron and Humboldt University, the TANGO Consortium of Taiwan, the University of Wisconsin at Milwaukee, Trinity College Dublin, Lawrence Livermore National Laboratories, IN2P3, University of Warwick, Ruhr University Bochum, Northwestern University and former partners the University of Washington, Los Alamos National Laboratories, and Lawrence Berkeley National Laboratories. Operations are conducted by COO, IPAC, and UW.

This research has made use of the SIMBAD database and the VizieR catalogue access tool, operated at CDS, Strasbourg, France. This research made use of Astropy,\footnote{\url{http://www.astropy.org}} a community-developed core Python package for Astronomy \citep{astropy2013, astropy2018}.

\facility{PO:1.2m}

\software{{\tt astropy} \citep{astropy2013,astropy2018}, {\tt dustmaps} \citep{green2018}, {\tt Matplotlib} \citep{hunter2007},  {\tt NumPy} \citep{harris2020}, {\tt SciPy} \citep{virtanen2020}.}



\end{document}